\documentclass[a4paper]{article}

\usepackage[latin1]{inputenc}
\usepackage{amsmath}
\usepackage{mathrsfs}
\usepackage{amsthm}
\usepackage{amsfonts}
\usepackage{empheq}
\usepackage{pifont}
\usepackage[english]{babel}
\usepackage[pdftex]{graphicx}

\usepackage{oldgerm}

\usepackage{setspace}

\usepackage{nicefrac}

\usepackage[T1]{fontenc}
\usepackage{fancybox}
\usepackage{float}
\usepackage{amssymb}

\usepackage{listings}

\newcommand{\xfrac}[2]{\mbox{\raisebox{0.4ex}{\ensuremath{\displaystyle #1}\hspace{-0.1ex}}\raisebox{-0.1ex}{\Large /}\hspace{-0.1ex}\raisebox{-0.2ex}{\ensuremath{\displaystyle #2}}}}

\newtheorem{theorem}{Theorem}[section]
\newtheorem{proposition}[theorem]{Proposition}
\newtheorem{lemma}[theorem]{Lemma}
\newtheorem{corollary}[theorem]{Corollary}
\theoremstyle{definition}
\newtheorem{remark}[theorem]{Remark}
\newtheorem{example}[theorem]{Example}

\newtheorem{algorithm}[theorem]{Algorithm}

\makeatletter
\g@addto@macro{\thm@space@setup}{\thm@headpunct{}}
\makeatother

\date{ }

\title{Aperiodic logarithmic signatures }
\author{Barbara Baumeister and Jan-Hendrik de Wiljes}

\begin{document}

\maketitle
\begin{abstract}
In this paper we propose a method to construct logarithmic signatures
which are not amalgamated transversal and further
do not even have a periodic block. The latter property
was crucial for the successful attack on the system 
\textit{MST}$_3$  by Blackburn et al. \cite{cid1}.
The idea for our construction is based on the theory in Szab\'o's book
about group factorizations \cite{sza1}.
\end{abstract}
\section{Introduction}

In the 80's Magliveras, Stinson and van Trung introduced two public key cryptosystems, \textit{MST}$_1$ and \textit{MST}$_2$, based on factorizations, covers and logarithmic signatures, of finite nonabelian groups \cite{mag1}. Recently, Lempken, Magliveras, van Trung and Wei \cite{lem1} developed a third cryptosystem, \textit{MST}$_3$.

A main question is how to produce covers and logarithmic signatures for a group. Blackburn et al. \cite{cid1} suggested a construction of so called amalgamated transversal logarithmic signatures from exact transversal logarithmic signatures (for the definition see Section \ref{exacttransversalLS}). Based on the use of these amalgamated transversal logarithmic signatures they presented a successful attack on the system \textit{MST}$_3$.

In this paper we propose a method to construct logarithmic signatures which are not amalgamated transversal and further do not even have the property of being periodic, which was crucial for breaking the system \textit{MST}$_3$ (see cases $2$ and $3$ in subsection $4.3$ in \cite{cid1}). The idea for this construction is based on the theory in Szab\'o's book about group factorizations \cite{sza1}.

The paper is organized as follows: In Section \ref{Covers_And_LogarithmicSignatures} covers and logarithmic signatures will be introduced and some basic facts will be presented.  We shortly introduce the cryptosystem \textit{MST}$_3$, for further information see also \cite{lem1} or \cite{cid1}. Then we introduce the in \cite{lem1}  proposed platform groups, the Suzuki $2$-groups.  The question of how to construct logarithmic signatures will be the main issue of Section \ref{some_classes_of_logarithmic_signatures}. In Section \ref{ATAbschnitt} we present the method for the construction of aperiodic logarithmic signatures. We will close with some final thoughts and remarks on further research in Section \ref{conclusion}.

\section{Covers and logarithmic signatures}\label{Covers_And_LogarithmicSignatures}

The cryptosystem \textit{MST}$_3$ is based on the use of covers and logarithmic signatures.
We will introduce them in this section and give a short overview of necessary results. Further information can be found in \cite{cus}, \cite{lem1}, \cite{lem2}, \cite{mag0} and \cite{mag1}. Throughout this paper, $G$ denotes a finite group and every set is assumed to be finite.

Let $K\subseteq G$ and $\alpha = [A_1,\dots,A_s]$ be a sequence of sequences $A_i=[a_{i,1},\dots,a_{i,r_i}]$ with $a_{i,j}\in G$, such that $\sum\limits_{i=1}^s{|A_i|}$ is bounded by a polynomial in $\lceil\log{|K|}\rceil$. Then $\alpha$ is a \textit{cover  for $K\subseteq G$}, if every product $a_{1,j_1}\cdots a_{s,j_s}$ lies in K and if every $g\in K$ can be written as
\begin{equation}\label{elementfactorization}
g = a_{1,j_1}\cdots a_{s,j_s}
\end{equation}
with $j_i\in\{1,\dots,|A_i|\}$.
We denote the set of all covers for $K\subseteq G$ by $\mathcal{C}(K|G)$. If, moreover, the tuple $(j_1,\dots,j_s)$ is unique for every $k\in K$ then $\alpha$ is called a \textit{logarithmic signature for $K$}. The set of all logarithmic signatures for $K$ is denoted by $\Lambda(K|G)$.

We call the product $a_{1,j_1}\cdots a_{s,j_s}$ in (\ref{elementfactorization}) a \textit{factorization} of $g$ w.r.t. $\alpha$. Two factorizations $a_{1,j_1}\cdots a_{s,j_s}$ and $a_{1,h_1}\cdots a_{s,h_s}$ of $g$ are \textit{different} if $(j_1,\dots,j_s)\neq(h_1,\dots,h_s)$. (Note that for $\alpha=[[a,a],[b,b]]$ the element $ab$ has four different factorizations $a\cdot b$.)

If $\alpha = [A_1,\dots,A_s]\in\mathcal{C}(K|G)$ with $r_i := |A_i|$ for all $i\in\{1,\dots,s\}$, then the sequence $A_i$ is called a \textit{block of} $\alpha$ and the sequence $(r_1,\dots,r_s)$  the \textit{type of} $\alpha$. The \textit{length of $\alpha$} is
\[l(\alpha) := \sum\limits_{i=1}^s{r_i}.
\]
Covers of minimal length are noteworthy due to the fact that less memory capacity has to be used. The interested reader is referred to \cite{lem2} for information on this issue.

For the application in cryptography the following distinction is made. A logarithmic signature $\beta\in\Lambda(K|G)$ is \textit{tame} if every $g\in K$ can be factorized polynomial in $\lceil log|K|\rceil$ w.r.t. to $\beta$, otherwise $\beta$ is called \textit{wild}.

The following map $\breve{\alpha}$ is used during the encryption and decryption procedure of the cryptosystem \textit{MST}$_3$. Later on we will identify factorizing w.\,r.\,t. a cover $\alpha$ with inverting $\breve{\alpha}$.

Let  $\alpha = [A_1,\dots,A_s]\in\mathcal{C}(K|G)$ be a cover for $K\subseteq G$ of type $(r_1,\dots,r_s)$ with $A_i = [a_{i,1},\dots,a_{i,r_i}]$ and let
\[
m := \prod\limits_{i=1}^sr_i\text{, }\, m_1 := 1\text{ and }m_i := \prod\limits_{l=1}^{i-1}r_l\text{ for all }i\in\{2,\dots,s\}.
\]
Let $\tau_\alpha$ be the canonic bijection from $\mathbb{Z}_{r_1}\times\cdots\times\mathbb{Z}_{r_s}$ to $\mathbb{Z}_m$, i.\,e.
\[
\tau_\alpha : \mathbb{Z}_{r_1}\times\cdots\times\mathbb{Z}_{r_s}\rightarrow\mathbb{Z}_m,
\tau_\alpha(j_1,\dots,j_s) := \sum\limits_{i=1}^sj_im_i.
\]
That is a generalization of $n$-ary representations.
Let $\breve{\alpha}: \mathbb{Z}_m\rightarrow K$ be the surjection:
\[
\breve{\alpha}(x) := a_{1,j_1+1}\cdots a_{s,j_s+1},~\mbox{where}~(j_1,\dots,j_s) = \tau_\alpha^{-1}(x).
\]

Note that $\tau_\alpha^{-1}$ can be computed efficiently (using Euclid's algorithm) and therefore the same is true for $\breve{\alpha}$. Moreover, the map $\tau_\alpha$ does only depend on the type of $\alpha$, i.\,e. for $\alpha,\beta\in\mathcal{C}(K|G)$ we have
\[
\tau_\alpha = \tau_\beta\Leftrightarrow\alpha\text{ and }\beta\text{ are of the same type}.
\]

Let $g\in G$ and let $\alpha_g$ be the number of pairwise different factorizations $a_{1,j_1}\cdots a_{s,j_s}$ of $g$ w.r.t. $\alpha$. Then $g$ has exactly $\alpha_g$ different preimages w.\,r.\,t. $\breve{\alpha}\circ\tau_\alpha$, namely the tuples $(j_1,\dots,j_s)$ with $g=a_{1,j_1}\cdots a_{s,j_s}$. That is the connection to equation (\ref{elementfactorization}). Therefore, a logarithmic signature $\beta\in\Lambda(K|G)$ is tame if we can compute $\breve{\beta}^{-1}$ polynomial in $\lceil log|K|\rceil$.

For $\mathcal{F}\in\{\mathcal{C},\Lambda,\dots\}$ we use the notation
\[
\mathcal{F}(G):=\mathcal{F}(G|G).
\]

\section{The cryptosystem MST$_3$}\label{MST_3}

Alice chooses a public non-abelian group $G$ with large center $Z$ and generates
\begin{itemize}
  \item a tame logarithmic signature $\beta = [B_1,\dots,B_s]\in\Lambda(Z)$ of type $(r_1,\dots,r_s)$
  \item and a random cover $\alpha = [A_1,\dots,A_s]\in\mathcal{C}(K|G)$ for a subset $K$ of $G$
 with $a_{i,j_i}\in G\backslash Z$ for all $i\in\{1,\dots,s\}$ and $j_i\in\{1,\dots,r_i\}$, which is of the same type as $\beta$.
\end{itemize}
Then she chooses random elements $t_0,\dots,t_s\in G\backslash Z$ and computes the following covers:
\begin{itemize}
  \item $\tilde{\alpha} = [\tilde{A}_1,\dots,\tilde{A}_s]$, whereat $\tilde{A}_i = t_{i-1}^{-1}A_it_i$ for all $i\in\{1,\dots,s\}$,
  \item $\gamma := [H_1,\dots,H_s]$ with $H_i:=[b_{i,1}\tilde{a}_{i,1},\dots,b_{i,r_i}\tilde{a}_{i,r_i}]$ for all $i\in\{1,\dots,s\}$.
\end{itemize}
The public key is $(\alpha, \gamma)$ and the private key is $(\beta,t_0,\dots,t_s)$.

To encrypt an element $x\in\mathbb{Z}_{|Z|}$, Bob computes
$y_1 = \breve{\alpha}(x)$ and $y_2 = \breve{\gamma}(x)$ and  sends $y = (y_1,y_2)$ to Alice.

Alice decrypts $y$ by calculating 
$\breve{\beta}^{-1}(y_2t_s^{-1}y_1^{-1}t_0)$ which equals $x$.
As $\beta$ is tame, the decryption-algorithm is efficient.

The cryptographic hypothesis is the problem of factorizing w.\,r.\,t. the random cover $\alpha$. Furthermore it has to be hard for the attacker to reconstruct the private key by using the public key. For information on these two issues we refer the reader to \cite{cid1}, \cite{vas3} and \cite{mag3}.

\begin{remark}Lempken, Magliveras, van Trung and Wei \cite{lem1} demand two additional properties.

Firstly the group $G$ should not be a direct product of $Z$ and a subgroup $U\leq G$, otherwise the system could be weakened using Schreier-trees \cite{lem1}. 

The second assumption is $a_{i,j}a_{i,l}^{-1}\notin Z$ for all $i\in\{1,\dots,s\}$ and $j\neq l$. However, Blackburn et al. \cite{cid1} didn't use that property for their attacks, because it holds for a large number of public keys and it is not required during the encryption and decryption process.
\end{remark}

Lempken et al. \cite{lem1} suggested the use of Suzuki $2$-groups (see also \cite{hup2} and \cite{hup3}) as platform-groups for the system:

Let $\theta\neq id$ be an odd order field automorphism of $\mathbb{F}_q$ ($q=2^n$). We then define the Suzuki $2$-group as
\[
G := \left\{S(c,d) : c,d\in\mathbb{F}_q\right\},
\]
where
\[
S(c,d) := \begin{pmatrix}
   1  &  0  &  0  \\
   c  &  1  &  0  \\
   d  &  c^{\theta}  &  1  \\
\end{pmatrix}.
\]
\begin{lemma}\label{center_lemma}
The center $Z(G)=\{S(0,d) : d\in\mathbb{F}_q\}$ is an elementary abelian $2$-group.
\end{lemma}

We will now concentrate on the construction of $\beta$ and we will restrict us, motivated by the Lemma \ref{center_lemma}, to elementary abelian $2$-groups, although all results in Section \ref{ATAbschnitt} hold for every abelian group.

\section{Classes of logarithmic signatures}\label{some_classes_of_logarithmic_signatures}

\subsection{Exact transversal logarithmic signatures}\label{exacttransversalLS} 

A logarithmic signature $\beta=[B_1,\dots,B_s]$ for a group $G$ is called \textit{l-exact transversal} (\textit{r-exact transversal}) if there is a subgroup chain
\[
G = G_0 > G_1 > \dots > G_s = \{1\},
\] 
such that $B_i$ is a left (right) transversal of $G_i$ in $G_{i-1}$ for all $i\in\{1,\dots,s\}$. A logarithmic signature is said to be \textit{exact transversal} if it is l-exact transversal or r-exact transversal. We denote the set of all exact transversal logarithmic signatures for a group $G$ by $\mathcal{ET}(G)$.

\begin{remark}
The block $B_s$ of an exact transversal logarithmic signature $\beta$ is a subgroup of $G$, more precisely $B_s=G_{s-1}$. Moreover, $[B_i,\dots,B_s]$ is an exact transversal logarithmic signature for $G_{i-1}$.
\end{remark}

\subsection{Amalgamated transversal logarithmic signatures}\label{ATLS}

Let $\beta=[B_1,\dots,B_s]$ be an exact transversal logarithmic signature of type $(r_1,\dots,r_s)$ for an abelian group $G$.
Blackburn et al. \cite{cid1} define the following operations on $\beta$:
\begin{itemize}
  \item permute elements within each $B_i$,
  \item permute the $B_i$,
  \item replace $B_i$ by a translate $B_ig$ for some $g\in G$,
  \item amalgamate two sets $B_i$ and $B_j$ by the single set $B_i \cdot B_j:= \{gh~|~g \in B_i, h \in B_j\}$.
\end{itemize} 
The logarithmic signatures that are constructed from an exact transversal logarithmic signature by applying a finite number of the four previous maps 
are called \textit{amalgamated transversal} logarithmic signatures,  see \cite{cid1}. We will denote the set of amalgamated transversal logarithmic signatures for a group $G$ by $ \mathcal{AT}(G)$.

The amalgamated transversal logarithmic signatures  have the special property of being periodic, which Blackburn et al. \cite{cid1} used to break \textit{MST}$_3$ under the assumption that the platform-group $G$ is a Suzuki-$2$-group. 
A subset $B$ of an abelian group $G$ is called \textit{periodic} if there exists a $g\in G\backslash\{1\}$ (the \textit{period}) with $gB = B$. Let
$P(B):=\{g\in G\backslash\{1\} : gB = B\}
$ be the \textit{set of periods} of $B$.

\begin{proposition}[Blackburn et al. \cite{cid1}, Lemma 2.1]\label{ATperiodischLemma}
Let $G$ be an abelian group and $\beta\in\mathcal{AT}(G)$. Then at least one of the blocks $B_i$ of $\beta$ is periodic.
\end{proposition}

Blackburn et al showed that every amalgamated transversal  logarithmic signature can be used in \textit{MST}$_3$. Their proof is based on
Proposition~\ref{ATperiodischLemma},  see \cite{cid1}.

\begin{theorem}[Blackburn et al. \cite{cid1}, Lemma 2.2]
Let $G$ be an elementary abelian $2$-group. Every logarithmic signature $\beta\in\mathcal{AT}(G)$ is tame.
\end{theorem}

\section{Constructing aperiodic tame logarithmic signatures}\label{ATAbschnitt}

Since the usage of amalgamated transversal logarithmic signatures leaves the cryptosystem insecure, we are in need to find new ways of constructing tame logarithmic signatures, preferably some without periodic blocks.  In this section we introduce an algorithm to construct tame logarithmic 
signatures without periodic blocks.

As in a logarithmic signature $\beta$ every group element is at most once in a block and as the position of the element inside a block is irrelevant for the tameness of $\beta$, see Theorem 4.4, we will consider sets instead of sequences.

We call a logarithmic signature $\beta\in\Lambda(G)$ \textit{aperiodic} if non of the blocks $B_i$ is periodic. The set of all aperiodic logarithmic signatures for a group $G$ is denoted by $\mathcal{A}(G)$.

\begin{theorem}[Szab\'o \cite{sza1}, Theorem 7.3.1]\label{SzaboTheorem}
Let $G$ be an elementary abelian $2$-group. There exists an aperiodic logarithmic signature $\beta$ of type $(r_1,\dots,r_s)$ with $r_1\geq\dots\geq r_s\geq 2$ if
\begin{itemize}
  \item $s = 2$ and $r_2\geq 8$ or
  \item $s\geq 3$ and $r_1\geq 8$, $ r_s\geq 4$ holds.
\end{itemize}
There does not exist an aperiodic logarithmic signature of type $(r_1,\dots,r_s)$  with $r_1\geq\dots\geq r_s\geq 2$
if one of the following cases holds:
\begin{itemize}
  \item $r_s = 2$,
  \item $s = 1$,
  \item $s = 2$ and $r_2|4$,
  \item $s\geq 3$ and $r_1|4,\dots,r_s|4$.
\end{itemize}
\end{theorem}

We are going to use the idea of the proof  of this theorem to construct tame aperiodic logarithmic signatures for elementary abelian $2$-groups, for example for the center of a Suzuki $2$-Group.

\subsection{The algorithm}

Now we are presenting the algorithm which constructs a new logarithmic signature out of a subgroup and a left transversal of that subgroup. The realization of some rather vague steps in the algorithm, namely the construction of $\delta$ and all $\alpha^{(j_1,\dots,j_s)}$,  will be discussed in the last part of the paper.

\begin{algorithm}\label{algorithm}

We start with an abelian group $G$, choose a subgroup $U$ of $G$ and a transversal $R$ of $U$ in $G$. 
Then we generate
\[
\delta = (D_1,\dots,D_s)\in\Lambda(R)
\]
with
\[
D_i = \{d_{i,1},\dots,d_{i,r_i}\}
\]
of type $(r_1,\dots,r_s)$ and logarithmic signatures
\[
\alpha^{(j_1,\dots,j_s)} := \left(A_1^{(j_1)},\dots,A_s^{(j_s)}\right)\in\Lambda(U)
\]
for all $(j_1,\dots,j_s)\in\{1,\dots,r_1\}\times\dots\times\{1,\dots,r_s\}$.
We get $\beta := (B_1,\dots,B_s)$ by
\begin{align*}
B_1& := d_{1,1}A_1^{(1)}\cup\dots\cup d_{1,r_1}A_1^{(r_1)}, \cdots, \\
B_s &:= d_{s,1}A_s^{(1)}\cup\dots\cup d_{s,r_s}A_s^{(r_s)}.
\end{align*}

\end{algorithm}

Notice that we needed all the logarithmic signatures  $\alpha^{(j_1,\dots,j_s)}$ to be able to produce
an aperiodic  logarithmic signature.

\begin{example}\label{ExampleForExponenentSix}
We choose $G=\langle u,v,w,x,y,z\rangle=2^6$, $U = \langle u,v,w,x\rangle$, $R=\{1,y,z,yz\}$ and set
\[
D_1 :=\{1,z\}\text{, }D_2 := \{1,y\}.
\]
and
\begin{align*}
A_1^{(1)} & := \{1,u,v,uv\}, A_1^{(2)} := \{1,w,x,wx\},\\
A_2^{(1)} & := \{1,uw,vx,uvwx\}, 
A_2^{(2)} := \{1,ux,uvw,vwx\}.
\end{align*}
We get
\begin{align*}
B_1 & :=\{1,u,v,uv,z,wz,xz,wxz\},
B_2  :=\{1,uw,vx,uvwx,y,uxy,uvwy,vwxy\}.
\end{align*}
Neither of these two blocks is periodic. It follows that $\beta\in\mathcal{A}(G)$ of type $(8,8)$.
\end{example}

\begin{theorem}\label{keinATsatz}
The sequence $\beta$  constructed by the algorithm \ref{algorithm} is a logarithmic signature for $G$ of type $(l_1, \ldots,l_s)$, where
$l_i = \sum_{j = 1}^{r_i}|A_i^{(j)}|$.
\end{theorem}

We denote a logarithmic signature which can be obtained from $U$ and $R$ by the construction
above \textit{decomposed and reunited} out of $U$ and $R$, shortly  \textit{d.r.}, and we denote
the set of logarithmic signatures for a group $G$ which are d.r. by $\mathcal{DR}_G(U,R,\mathcal{E}(U|G),\mathcal{F}(R|G))$ where $\mathcal{E},\mathcal{F}\in\{\Lambda,\mathcal{ET},\mathcal{AT},\dots\}$.

\begin{remark}
Every logarithmic signature $\beta=(B_1,\dots,B_s)\in\Lambda(G)$ is d.r. out of $U = G$ and $R = \{1\}$: Set
 $\delta=(1,\dots,1)$ and $A_i^{(1)}=B_i$ for all $i=1,\dots,s$.
\end{remark}

An immediate question is how the choice of $U$ and $R$ influences the set $\mathcal{DR}_G(U,R,\mathcal{E}(U|G),\mathcal{F}(R|G))$.
Another question is which logarithmic signatures are constructible out of the pair $(U,R)$  when we choose $\gamma$ and $\alpha^{(j_1,\dots,j_s)}$ to be for example exact transversal only. 
  
It is possible to construct an aperiodic logarithmic signature by using only \textit{total exact transversals}, i.e. exact transversals where every block is a subgroup, see Example~\ref{ExampleForExponenentSix} above.

\begin{proposition}
A logarithmic signature which is d.r. is tame if $\delta$ and all $\alpha^{(j_1,\dots,j_s)}$ are tame and if for every $g\in G$ the coset representative in $R$ which lies in the same coset as $g$ can be found efficiently.  
\end{proposition}

\subsection{Aperiodicity of  $\beta$ }

From now on we assume that $\beta = (B_1, \ldots , B_s)$ is constructed by the algorithm~\ref{algorithm} and we use the notation introduced there. Next we summarize some basic facts. After that we show how to choose the sets $A_i^{(j)}$ to force the non-periodicity of $B_i$.

\begin{lemma}\label{RepraesentantenVerschiedenLemma}
We have  $d_{i,j}^{-1}d_{i,k}\notin U$ for all $i=1,\dots,s$ and $j,k=1,\dots,r_i$ with $j\neq k$.
\end{lemma}

\begin{proof}
We assume that there are $i$ and $j\neq k$ with $d_{i,j}^{-1}d_{i,k}\in U$. We consider the two factorizations 
$$
d_{1,1}\cdots d_{i-1,1}d_{i,j}d_{i+1,1}\cdots d_{s,1}\quad\text{and}\quad d_{1,1}\cdots d_{i-1,1}d_{i,k}d_{i+1,1}\cdots d_{s,1}
.$$
 These elements of $R$ are in different cosets of $U$ in $G$. On the other hand we have\\
$
(d_{1,1}\cdots d_{i-1,1}d_{i,j}d_{i+1,1}\cdots d_{s,1})^{-1}d_{1,1}\cdots d_{i-1,1}d_{i,k}d_{i+1,1}\cdots d_{s,1} = d_{i,j}^{-1}d_{i,k}\in U,
$
which is not possible.
\end{proof}

\begin{lemma}\label{UntergruppenGleichheit}
Let $A,B\leq G$. Then $A=B$ if and only if there exists an element $g\in G$ with $gA = B$.
\end{lemma}

\begin{lemma}\label{PeriodeBildetAufGanzesAbLemma}
If $B_i$ is periodic with period $g\in G$, then for every $d_{i,j}A_i^{(j)}$ there is a $k\in\{1,\dots,r_i\}$, such that
\[
gd_{i,j}A_i^{(j)} = d_{i,k}A_i^{(k)}.
\] If additionally $A_i^{(j)},A_i^{(k)}\leq G$ holds, then 
$A_i^{(j)} = A_i^{(k)}$.
\end{lemma}

\begin{proof}
Assume there is no such $k$. Then we have $a_1,a_2\in A_i^{(j)}$ with $a_1\neq a_2$ and $b\in A_i^{(e)}$, $c\in A_i^{(l)}$ for $e\neq l$, such that
$
gd_{i,j}a_1 = d_{i,e}b\text{ }\text{ and }\text{ }gd_{i,j}a_2 = d_{i,l}c
$.
From that it follows
$
d_{i,l}^{-1}d_{i,e}^{} = ca_2^{-1}a_1^{}b^{-1}\in U,$
which  is a contradiction to Lemma \ref{RepraesentantenVerschiedenLemma}. This shows the first statement.
The second part follows from Lemma \ref{UntergruppenGleichheit}.
\end{proof}

To describe the periodic signatures $\beta$ we introduce
for  $i\in\{1,\dots,s\}$ the  set
\[
D_i^{(j)} := \left\{d_{i,k} : A_i^{(k)} = A_i^{(j)}\right\}
\]
of elements $d_{i,k}$ that have the same corresponding subset $A_i^{(k)}$.
Then we immediately obtain the following:

\begin{lemma}\label{periodischLemma}
$B_i$ is periodic if one of the following holds:
\begin{itemize}
\item[(i)]
$
\bigcap\limits_{j=1}^{r_i}P\left(A_i^{(j)}\right)\neq\emptyset.
$
\item[(ii)]  $\bigcap\limits_{j=1}^{r_i}P\left(D_i^{(j)}\right)\neq\emptyset.$
\end{itemize}
\end{lemma}

The special case $r_i=2$ or $3$ and pairwise different subgroups $A_i^{(j)}$ of the following theorem was proven in cooperation with Anja Nuss \cite{nuss}.

\begin{theorem}\label{PeriodizitaetsGenauDannWennSatz}
Let $A_i^{(j)}\leq G$ for all $j\in\{1,\dots,r_i\}$. Then $B_i$ is periodic if and only if
\[
\bigcap\limits_{j=1}^{r_i}P\left(X^{(j)}\right)\neq\emptyset
\]
holds for at least one $r_i$-tuple $\left(X^{(1)},\dots,X^{(r_i)}\right)\in\left\{A_i^{(1)},D_i^{(1)}\right\}\times\cdots\times\left\{A_i^{(r_i)},D_i^{(r_i)}\right\}$.
\end{theorem}

\begin{proof}
One part of the equivalence follows from Lemmas \ref{PeriodeBildetAufGanzesAbLemma} and  \ref{periodischLemma}.

Now assume that  $B_i$ is periodic. Let $g\in G$ be a period of $B_i$.  By Lemma \ref{PeriodeBildetAufGanzesAbLemma} we have
for every $j$ and $D_i^{(j)} = \{d_{i,{j_1}},\dots,d_{i,{j_k}}\}$ that
\[
g\left(d_{i,{j_1}}A_i^{(j)}\cup\dots\cup d_{i,{j_k}}A_i^{(j)}\right) = d_{i,{j_1}}A_i^{(j)}\cup\dots\cup d_{i,{j_k}}A_i^{(j)}.
\]
Moreover, every $d_{i,{j_l}}A_i^{(j)}$ is mapped to a $d_{i,{j_c}}A_i^{(j)}$ by multiplication with $g$. Therefore $g$ must be either an element of $A_i^{(j)}$, i.\,e. $g\in P\left(A_i^{(j)}\right)$ or $g$ permutes the elements of $D_i^{(j)}$, i.\,e. $g\in P\left(D_i^{(j)}\right)$.
\end{proof}

An immediate consequence is the  following equivalence.

\begin{corollary}\label{corollarySubgroupsDifferent}
Let $A_i^{(j)}\leq G$ for all $j\in\{1,\dots,r_i\}$ and $A_i^{(j)}\neq A_i^{(k)}$ for all $j,k\in\{1,\dots,r_i\}$ with $j\neq k$. Then $B_i$ is periodic if and only if
\[
\bigcap\limits_{j=1}^{r_i}P\left(A_i^{(j)}\right)\neq\emptyset.
\]
\end{corollary}
\hfill \qed

If at least one $A_i^{(j)}$ is not a subgroup of $G$, then the statement of Theorem \ref{PeriodizitaetsGenauDannWennSatz} does not hold 
anymore. The following example shows that we can already get a periodic block when $r_i = 2$.

\begin{example}\label{example_6}
We choose $G := \langle u,v,w,x,y,z\rangle = 2^6, U = \langle u,v,w,x\rangle$ and set
\begin{align*}
A_1^{(1)} := \{1,u,v,uvw\}\text{,} ~
A_1^{(2)}  := \{u,1,uv,vw\} = u^{-1}A_1^{(1)}
\end{align*}
and
\[
D_1 := \{1,y\}.
\]
Then we get
\[
B_1 = d_{1,1}A_1^{(1)}\cup d_{1,2}A_1^{(2)} = \{1,u,v,uvw,uy,y,uvy,vwy\},
\]
which has the period $uy$. But the other conditions of Theorem \ref{PeriodizitaetsGenauDannWennSatz} are fulfilled because of
\[
P\left(D_1^{(1)}\right) = P\left(D_1^{(2)}\right) = P\left(A_1^{(1)}\right) = P\left(A_1^{(2)}\right) = \emptyset.
\]
If  we set
\[
A_2^{(1)} := A_2^{(2)} := \{1,w,x,wx\}
~\mbox{and}~
D_2 := \{1,z\},
\]
then we get a logarithmic signature for $G$.
\end{example}

Next we  generalize Theorem \ref{PeriodizitaetsGenauDannWennSatz}.
For $G$  a group and $A,B\subseteq G$ we say that 
$A$ is a \textit{multiple} of $B$ if there is a $g\in G$ with $gA = B$.
Notice if $B$ is a subgroup of $G$ and $A$ a multiple of $B$, then $A$ is a left coset of $B$ in $G$.
We say that a multiple $A$ of $B$ is \textit{proper}, if $A \neq B$.

\begin{lemma}\label{PeriodeBildetAufGanzesAbLemma2}
If $B_i$ is periodic with period $g\in G$,  if $A_i^{(j)}$ is not a proper multiple of $A_i^{(k)}$ and if $gd_{i,j}A_i^{(j)}=d_{i,k}A_i^{(k)}$, then
$A_i^{(j)}=A_i^{(k)}$.
\end{lemma}

\begin{proof}
That follows immediately from the previous definition, because of $d_{i,k}^{-1}gd_{i,j}A_i^{(j)}=A_i^{(k)}$.
\end{proof}

\begin{theorem}\label{PeriodizitaetsGenauDannWennSatzVerallgemeinerung}
Suppose that  $A_i^{(j)}$ is not a proper multiple of $A_i^{(k)}$ for all $j,k\in\{1,\dots,r_i\}$. Then $B_i$ is periodic if and only if
\[
\bigcap\limits_{j=1}^{r_i}P\left(X^{(j)}\right)\neq\emptyset
\]
 for at least one $r_i$-tuple $\left(X^{(1)},\dots,X^{(r_i)}\right)\in\left\{A_i^{(1)},D_i^{(1)}\right\}\times\cdots\times\left\{A_i^{(r_i)},D_i^{(r_i)}\right\}$.
\end{theorem}

\begin{proof}
The proof is analog to the one of Theorem \ref{PeriodizitaetsGenauDannWennSatz} but we have to use Lemma \ref{PeriodeBildetAufGanzesAbLemma2} instead of Lemma \ref{PeriodeBildetAufGanzesAbLemma}.
\end{proof}

\begin{corollary}\label{CoroPeriodizitaetsGenauDannWennSatzVerallgemeinerung}
Suppose that  $A_i^{(j)}$ is not a multiple of $A_i^{(k)}$ for all $j,k\in\{1,\dots,r_i\}$. Then $B_i$ is periodic if and only if
\[
\bigcap\limits_{j=1}^{r_i}P\left(A_i^{(j)}\right)\neq\emptyset.
\]
\end{corollary}

\subsection{Concret construction for $G$ elementary abelian of order  $2^n$.}

We will construct aperiodic logarithmic signatures for elementary abelian $2$-groups $G$.
Such a logarithmic signature has already been constructed in Example \ref{ExampleForExponenentSix} for $G=2^6$.  Now we generate one for $G=2^7$ and then use these two logarithmic signatures to construct tame aperiodic logarithmic signatures for all groups $2^n$ with $n\geq 6$.

\begin{example}[see also Szab\'o \cite{sza1}, Theorem 7.3.1]\label{example_7}
We choose $G=\langle t,u,v,w,x,y,z\rangle=2^7$, $U = \langle u,v,w,x,y,z\rangle$, $R=\{1,t\}$ and set
\begin{align*}
A_1^{(1)} & := \{1,v,wx,vwx\}\text{,}~
A_1^{(2)}  := \{1,w,vz,vwz\},\\
A_2^{(1)} & := \{1,x,y,xyz\}\text{,}\\
A_3^{(1)} & := \{1,z,u,zuw\}.
\end{align*}
and
\[
D_1 := \{1,t\}\text{, }D_2 := \{1\}\text{, }D_3 := \{1\}.
\]
The resulting logarithmic signature $\beta$ is aperiodic of type $(8,4,4)$.
\end{example}

\noindent
{\bf General construction.}
Let $G=2^n$ be an elementary abelian group of order $n$ and let $\mathcal{B}=\{g_1,\dots,g_n\}$ be a generating set for $G$. We now decompose $G$ in the following way:
\[
G=\underbrace{U_1\times\dots\times U_s}_{=U}\times\underbrace{D_1\times\dots\times D_s}_{=R}
\]
where $U_1\times D_1$ is a small group with a known aperiodic logarithmic signature $\beta'$ (see Examples \ref{example_6} and \ref{example_7}) and $2\leq\left|D_i\right|<\prod_{j=1}^{i-1}\left|U_j\right|$ for $i\in\{2,\dots,s\}$. Then we choose for every $i\in\{2,\dots,s\}$ a subset $K_i:=\{k_i^{(1)},\dots,k_i^{(r_i)}\}\subseteq(U_1\times\dots\times U_{i-1})^\#$ of size $r_i:=|D_i|$. We construct the logarithmic signature $\beta=[\beta',B_2,\dots,B_s]$ using Algorithm~\ref{algorithm} by setting
\begin{align*}
\delta & :=  [D_2,\dots,D_s],\\
A_{i}^{(j)} & := \{1\} \cup \{k_i^{(j)} u:u \in U_i^\#\}\text{, for} ~i=2,\dots,s\text{ and }j=1,\ldots,r_i.
\end{align*}
Then no $A_i^{(j)}$ is the multiple of an $A_i^{(l)}$ for some $l \neq j$.
Therefore, Corollary \ref{CoroPeriodizitaetsGenauDannWennSatzVerallgemeinerung} implies that the resulting logarithmic signature $\beta$ for $G$ is aperiodic.

For security and storage issues it seems to be reasonable to choose small subgroups $U_i$ and $D_i$. Further, one should apply some of the operations from subsection \ref{ATLS} to $\beta$ to hide the subgroup $U_1\times D_1$, more precisely, the blocks of the logarithmic signature $\beta'$, otherwise an attacker could obtain a periodic (and therefore tame) logarithmic signature for $\xfrac{G}{(U_1\times D_1)}$.

If we want to store this logarithmic signature 
we are only in need to store a minimal generating set $\mathcal{B} = \cup_{l = 1}^{2s} \mathcal{B}_l$ of $G$
such that the subsets $\mathcal{B}_l$  generate $U_i$ and $D_i$, respectively,
and  the information which elements of $\mathcal{B}$ generate which subgroups $U_i$ and $D_i$.
The latter can be provided for example by a tuple $v\in\mathbb{Z}^{2s}$, and a strict total order on the $\mathbb{F}_2$-vector space $\mathbb{F}_2^n$, e.g. the lexicographical order, because the position of the elements is needed for the factorization.\\\\
\noindent
{\bf Factorization.} We define $V_i:=\sum_{k=1}^iv_k$ and use the following algorithm:
\begin{align*}
&\text{Let }y=\mathcal{K}_\mathcal{B}(g)\text{ be the coordinate vector of }g\text{ w.r.t. }\mathcal{B}\\
&\text{Let }j\in\mathbb{Z}^{s-1}\text{ be the tuple consisting only of ones}\\
&\text{for } i=s+1 \text{ to } 2s-1 \text{ do}\\
&\text{ }\quad\text{for } l=1+V_i \text{ to } V_{i+1} \text{ do}\\
&\text{ }\quad\quad\text{if } y_l= 0 \text{ then}\\
&\text{ }\quad\quad\quad j_{i-s}=j_{i-s}+2^{v_i-(l-V_i)}\\
&\text{Let }h\in\mathbb{Z}^{s-1}\text{ be the tuple consisting only of ones}\\
&\text{for } i=1 \text{ to } s-1 \text{ do}\\
&\text{ }\quad h_i=h_i+(j_i-1)2^{v_{i+1}}\\
&\text{ }\quad\text{for } l=1+V_i \text{ to } V_{i+1} \text{ do}\\
&\text{ }\quad\quad\text{if } y_l= 0 \text{ then}\\
&\text{ }\quad\quad\quad h_i=h_i+2^{v_i-(l-V_i)}
\end{align*}
Now we need to factorize the projection $y'$ of $y$ onto $U_1\times R_1$ which yields $x=\tau_{\beta'}^{-1}(g')$ where $y'=\mathcal{K}_{\mathcal{B}'}(g')$ and $\mathcal{B}'\subseteq\mathcal{B}$ a generating set of $U_1\times R_1$.

Altogether we get $\tau_\beta^{-1}(g)=(x,h_1,\dots,h_{s-1})$ and from that we receive $\breve{\beta}^{-1}(g)$.

Note that we have to treat $\beta'$ differently, but since $U_1\times D_1$ is small, we get the requested element in the factorization efficiently (meaning in $O(log_2|G|)$) by an exhaustive search.\\

\noindent
{\bf Complexity.}
Under the assumption, that comparison and arithmetic in $\mathbb{Z}$ can be done in $O(1)$, we can compute the complexity of the factorization-algorithm in the following way (worst case):
\begin{align*}
&\sum\limits_{i=s+1}^{2s-1}\sum\limits_{l=1+V_i}^{V_{i+1}}(v_i-l+V_i+4)+\sum\limits_{i=1}^{s-1}\left(v_{i+1}+3+\sum\limits_{l=1+V_i}^{V_{i+1}}(v_i-l+V_i+4)\right)\\
&=\sum\limits_{i=1}^{2s-1}\sum\limits_{l=1+V_i}^{V_{i+1}}(v_i-l+V_i+4)-\sum\limits_{l=1+V_s}^{V_{s+1}}(v_s-l+V_s+4)+\sum\limits_{i=1}^{s-1}(v_{i+1}+3)\\
&=\sum\limits_{i=1}^{2s-1}\left(v_{i+1}(v_i+V_i+4)-\sum\limits_{l=1+V_i}^{V_{i+1}}l\right)-\left(v_{s+1}(v_s+V_s+4)-\sum\limits_{l=1+V_s}^{V_{s+1}}l\right)+X\\
&=\sum\limits_{i=1}^{2s-1}\left(v_{i+1}(v_i+V_i+4)-\left(v_{i+1}V_i+\sum\limits_{l=1}^{v_{i+1}}l\right)\right)-Y+X\\
&=\sum\limits_{i=1}^{2s-1}\left(v_{i+1}\left(v_i-\frac{1}{2}v_{i+1}\right)\right)+\sum\limits_{i=1}^{2s-1}\left(\frac{7}{2}v_{i+1}\right)-Y+X\\
&\leq M\cdot n+\frac{7}{2}n-Y+X\qquad\qquad\qquad\qquad\qquad\left(\text{where }M=\max_i\left|v_i-\frac{1}{2}v_{i+1}\right|\right)
\end{align*}
with $X:=\sum\limits_{i=1}^{s-1}(v_{i+1}+3), Y:=\left(v_{s+1}(v_s+V_s+4)-\sum\limits_{l=1+V_s}^{V_{s+1}}l\right)$ and
\begin{align*}
-Y+X&=-\left(v_{s+1}v_s+4v_{s+1}-\left(\frac{1}{2}(v_{s+1})(v_{s+1}+1)\right)\right)+\sum\limits_{i=1}^{s-1}(v_{i+1}+3)\\
&=3s+\sum\limits_{i=1}^{s-1}v_{i+1}-\left(v_s+\frac{7}{2}-\frac{1}{2}v_{s+1}\right)v_{s+1}-3\\
&\leq 4n-\left(v_s+\frac{7}{2}-\frac{1}{2}v_{s+1}\right)v_{s+1}
\end{align*}
Since every $v_i$ is supposed to be small, especially $v_{s+1}$, $M$ will also be small and $v_s+\frac{7}{2}-\frac{1}{2}v_{s+1}$ will be nonnegative, independent of $n$. So in that case the runtime of the factorization-algorithm is $O(n)=O(log_2|G|)$ and, therefore, $\beta$ is tame.

\section{Conclusion}\label{conclusion}
We presented a new way to construct tame logarithmic signatures. The advantage of this method is the possibility to produce aperiodic logarithmic signatures which resist the attack proposed in \cite{cid1}. Although, one is in need to store $\delta$ and all $\alpha^{(j_1,\dots,j_s)}$ to factorize with respect to  $\beta$, this is also an aspect of security, because an attacker doesn't know those elements used during the construction but is in need to find them for being able to factorize w.r.t. $\beta$, as far as we know.

Further, we showed how to get a huge number of aperiodic tame logarithmic signatures by using the proposed algorithm. Although, those might not be enough, the fact that we mainly used exact transversal logarithmic signatures for the construction of our examples implies the assumption that many more aperiodic logarithmic signatures might be gained when using for example amalgamated transversal logarithmic signatures.

Still, it is not clear if the proposed algorithm has any weaknesses in view of the reconstruction of $\delta$ and the $\alpha^{(j_1,\dots,j_s)}$ from a given $\beta$ because of the known structure of the algorithm, although we conjecture that keeping the used generating set $\mathcal{B}$ a secret makes it hard to extract any information. Further, we don't know if $\beta$ is tame whether or not one knows $\delta$ and $\alpha^{(j_1,\dots,j_s)}$, which is also an important issue for an attacker.

\end{document}